\documentstyle{lamuphys}
\include{epsf}
\makeatletter
\let\chapter\hid@chapter
\makeatother
\begin{document}
\newcommand{\beq}{\begin{equation}}
\newcommand{\eeq}{\end{equation}}
\newcommand{\beqa}{\begin{eqnarray}}
\newcommand{\eeqa}{\end{eqnarray}}
\newcommand{\krig}[1]{\stackrel{\circ}{#1}}
\pagenumbering{arabic}
\title{New developments in threshold pion photo- and electroproduction}

\author{V\'eronique Bernard}

\institute{Laboratoire de Physique Th\'eorique,
F-67037 Strasbourg Cedex 2, France}

\maketitle

\begin{abstract}
Photoproduction of neutral and charged pions off nucleons and deuterium
has been precisely calculated in baryon chiral perturbation theory. I
review the predictions in light of the accurate data that have become available
over the last few years. Some progress in the description of neutral
pion electroproduction off protons is also discussed.
\end{abstract}
\section{Introduction}
With the advent of CW machines, pion production by real and virtual 
photons has become a major testing ground for predictions based on 
nucleon chiral dynamics. 
In particular over the last years there has
been considerable activity to precisely measure pion photo-and
electroproduction in the threshold region at various laboratories, like e.g.
MAMI, SAL and NIKHEF. 
On the other hand, refined calculations within heavy baryon chiral 
perturbation theory (HBCHPT) have been performed. Pion photo-and electroproduction
is not only interesting per se but also allows, as we will see in the
following, to get informations on other processes like for example
$\pi N$ scattering. In this respect the use of CHPT, see for example
\cite{leu:gass}, is particularly advantageous since it is a {\it
method} for solving QCD at low energy which links different
processes in a model--independent fashion thus allowing for a deeper
understanding of the underlying dynamics. Another very important
problem which can be addressed in dealing with these processes is 
isospin symmetry. It has always been one of the major goals in nuclear 
physics to understand isospin symmetry violation related to the light
quark mass difference $m_u-m_d$ and virtual photon effects. Although the light
quark mass ratio deviates strongly from unity, $m_d/m_u \sim 2$, see 
\cite{gass} and one could expect sizeable isospin violation, such effects
are effectively masked since $m_d-m_u  \ll
\Lambda$, with $\Lambda$ the scale of the strong interactions (which can be
chosen to be $4\pi F_\pi$ or 1~GeV or the mass of the $\rho$). To assess
the isospin violation through quark mass differences,
precise measurements and accurate calculations are mandatory. As pointed out
by  \cite{weinmass} long time ago, systems involving nucleons
can exhibit such effects to leading order in contrast to the suppression
in purely pionic  processes due to G--parity. Thus
pion photoproduction is well suited for such investigations since
a large body of (precise) data exists for various isospin channels.
   
Here I will report on the progress made since the MIT workshop, see
\cite{mit}, on the calculation of charged and neutral pion photo-and 
electroproduction near threshold in the framework of HBCHPT. In
this framework the nucleons are treated as very heavy static sources. 
HBCHPT is a triple expansion in external momenta, quark masses and 
inverse powers of the nucleon mass (collectively denoted by the small
parameter $q$).  Calculations which will be discussed here are ${\cal O}
(q^n)$ with n=4 for the S-wave and n=3 for the P-waves 
and only take into account isospin
breaking effects which are believed to be dominant, namely those
through the pion mass difference in the loops.
  
\section{Formal aspects}
\subsection{Effective field theory}
The effective Lagrangian which will be needed here
consists of the following pieces:
\begin{equation}
{\cal L}_{\rm eff} = {\cal L}_{\pi\pi}^{(2)} +{\cal L}_{\pi\pi}^{(4)} +
 {\cal L}_{\pi N}^{(1)}
 + {\cal L}_{\pi N}^{(2)}  + {\cal L}_{\pi N}^{(3)}
 + {\cal L}_{\pi N}^{(4)}  + {\cal L}_{N N}^{(0)}
 + {\cal L}_{N N}^{(1)} \,\, ,
\end{equation}
where the index $(i)$ gives the chiral dimension $d_i$ (number of derivative
and/or meson mass insertions).
${\cal L}_{\pi\pi}^{(2)}+ {\cal L}_{\pi N}^{(1)} $ is the non-linear 
$\sigma$ model Lagrangian coupled to nucleons.
The terms from  ${\cal L}_{\pi N}^{(3)} + {\cal L}_{\pi N}^{(4)}$ 
contributing to the single--nucleon photoproduction amplitudes 
are given in \cite{bkmz}. ${\cal L}_{N N}^{(1)}$ has been used 
in the nuclear force calculation of \cite{ubi2} and enters the calculation
of (neutral) pion photoproduction on deuterium by \cite{bblm}, which will be
discussed below. In HBCHPT the nucleon mass term is replaced by a string
of $1/m$ suppressed interactions so that 
${\cal L}_{\pi N}^{(n)}, n \ge 2$, contains $1/m$ terms as well as counterterms
whose coefficients are the famous low--energy constants
(LECs). Some of these counterterms cancel the
divergences of certain loop diagrams and thus are scale dependent,
$C_i =C_i^r(\mu) + \Gamma_i L(\mu)$ where $\mu$ is the scale of dimensional
regularization (naturally the physical amplitudes are scale independent).
In the
following these LECs which are not fixed by chiral symmetry,
will either be fitted to experimental data or will be obtained using the
principle of resonance saturation. In the meson sector it can indeed be 
shown that the numerical values of the renormalized LECs $L_i^r(\mu = M_\rho)$
can be understood to a high degree of accuracy from resonance saturation,
i.e. they can be expressed in terms of resonance masses and coupling 
constants of the low-lying vector ($V$), axial vector ($A$), scalar ($S$)
and pseudoscalar ($P$) multiplets, see \cite{egpr}. In the nucleon sector
there exists no proof of this principle,  however it seems to work rather
well in the case of  ${\cal L}_{\pi N}^{(2)}$ as has been demonstrated in 
\cite{bkma} and can be seen in table 1 which gives the values of the seven
finite LECs at that order. The first four have been determined (second column)
from a best fit to a set of nine subthreshold and threshold $\pi N$
observables that to one--loop
order $q^3$ are given entirely in terms of tree graphs including
insertions from these LECs and finite loop contributions,
but with none from the 24 new LECs of ${\cal L}_{\pi N}^{(3)}$.  The other three
can be determined from the
strong neutron--proton mass difference ($c_5$, which is only relevant in the
case $m_u \neq m_d$) and from the anomalous magnetic moments of the
proton and the neutron ($c_6$, $c_7$). Note that
the scalar mass to coupling constant
ratio $M_S / \sqrt{g_S}$ needed to saturate the LEC $c_1$ is in
perfect agreement with typical ratios obtained in boson--exchange
models of the NN force, where the $\sigma$--meson models the strong
pionic correlations coupled to nucleons. Note also that the important 
$\Delta$ degree of freedom is included in the determination of the LECs
through resonance exchange. I will come back to this point in the following. 

\renewcommand{\arraystretch}{1.3}

\begin{center}

\begin{tabular}{|l|r|r|r|c|}
    \hline
    $i$         & $c_i \quad \quad$   &  
                  $ c_i^{\rm Res} \,\,$ cv & 
                  $ c_i^{\rm Res} \,\,$ ranges &
                  Res \quad \quad    \\
    \hline
    1  &  $-0.93 \pm 0.10$  &  $-0.9^*$ & --  & $S$ \\
    2  &  $3.34  \pm 0.20$  &  $3.9\,\,$ & $2 \ldots 4$ & $\Delta, R$ \\    
    3  &  $-5.29 \pm 0.25$  &  $-5.3\,\,$ 
                                     & $-4.5 \ldots -5.3$ & $\Delta, R, S$ \\
    4  &  $3.63  \pm 0.10$   &  $3.7\,\,$ 
                                     & $3.1 \ldots 3.7$ & $\Delta, R, \rho$ \\    
    5  &  $-0.09 \pm 0.01$   & $-\,\,$ & $-\,\,$ &  \\    
    \hline
    6           &  $5.83$               & $6.1\,\,$  & $-\,\,$ & $\rho$  \\
    7           &  $-2.98$              & $-3.0\,\,$ & $-\,\,$ & $\rho, \omega $\\
    \hline
  \end{tabular}

\smallskip 

\end{center}

\noindent Table 1: Values of the LECs $c_i$ in GeV$^{-1}$
for $i=1,\ldots,5$. The LECs $c_{6,7}$ are dimensionless.
Also given are the central values (cv) and the ranges
for the $c_i$ from resonance
exchange.  The $^*$ denotes an input quantity. $R$ and $S$ denote the Roper
and the scalar resonances respectively

\subsection{Threshold pion photo-and electroproduction}
Consider the process $\gamma(k) + N(p_1) \to \pi^a(q) + N(p_2)$,
with $N$ denoting the nucleon (proton or neutron), $\gamma$ a real ($k^2= 0$) or
a virtual ($k^2 < 0$)
photon and $\pi^a$ a pion of isospin $a$. The 
polarization vector of the photon is
denoted by $\epsilon_\mu$. In the threshold region, the
three--momentum $\vec q$ of the pion is small and vanishes at
threshold. It is therefore  
advantageous to perform a multipole decomposition since at threshold only the
S--waves survive and close to threshold one can confine oneself to
S-- and P--waves. The corresponding multipoles are called
$(E, \, M, \, L)_{l \pm}$, where $E, \,
M, \, L$ stands for electric,  magnetic and longitudinal (this last one
of course only comes into play for virtual photons),
 $l = 0,1,2, \ldots$ the pion orbital angular momentum and
the $\pm$ refers to the total angular momentum of the pion-nucleon system, $j
=l \pm 1/2$. These multipoles parametrize the structure of the 
nucleon as probed with low energy photons.
Consequently, the T--matrix  depends on seven (only four survive for real 
photons) complex multipoles and takes the following form:
\begin{eqnarray}
&{m \over 4 \pi \sqrt s} \, \vec T \cdot \vec \epsilon =
 i \vec \sigma \cdot \vec \epsilon\, \biggl(E_{0+} + 
\hat q \cdot \hat k
\, P_1 \biggr) + i \vec \sigma \cdot \hat k \, \vec \epsilon \cdot \hat q \, 
P_2 + (\hat q \times \hat k)\cdot \vec \epsilon \, P_3  \nonumber \\ 
& + i \vec \sigma \cdot \hat k \, \vec \epsilon \cdot \hat k\, \biggl(L_{0+} - E_{0+} +
\hat q \cdot  \hat k\,(P_4 -P_5-P_1-P_2) \biggr) 
+ i  \vec \sigma \cdot \hat q \, \vec \epsilon \cdot \hat k \, P_5 \,
\label{defJ}
\end{eqnarray} 
The quantities $P_{1 - 5}$ represent the following combinations of the five
$P$-waves, $E_{1+}, \, M_{1+}$, $M_{1-}, \, L_{1+}$, \, $L_{1-}$,
\beqa 
P_1 = 3E_{1+} + M_{1+} &-& M_{1-} \, ,
  \,
P_2 = 3E_{1+} - M_{1+} + M_{1-} \, , \,
P_3 = 2 M_{1+} + M_{1-} \cr 
P_4 &=& 4 L_{1+} + L_{1-} \, , \,
P_5 = L_{1-} - 2 L_{1+} \,\, . 
\label{pw}
\eeqa 
All   
these amplitudes are easily calculable within CHPT. They have the 
conventional isospin decomposition (to first order in the electromagnetic
coupling),
\beqa
A(s,u) = A^{(+)}(s,u) \delta_{a,3} +  A^{(-)}(s,u) {1 \over 2} [\tau_a, \tau_3]
+A^{(0)}(s,u) \tau_3
\label{iso}
\eeqa
There are however, four physical channels, two charged reactions and two neutral 
ones which will be discussed next. Thus there exists a triangle relation
relating one of the physical amplitudes to the others. This relation should
of course only hold if isospin is an exact symmetry. It is then clear that
it is very important to have a very precise determination of the four
physical channels in order to measure isospin violation. 

\section{Charged Pion Photoproduction}
Charged pion photoproduction is a particularly interesting process since
it allows to

{$\bullet$} investigate the violation of isospin symmetry beyond leading order 
in electromagnetism.
For that one of course
needs to know extremely precisely the four physical photoproduction reactions as discussed
previously.

{$\bullet$} determine $\pi N$ scattering lengths. 

{$\bullet$} give a stringent constraint on the much discussed value of 
the pion-nucleon coupling constant $g_{\pi N}$ via the \cite{gmo}
sum rule combined with the Panofsky ratio.

\noindent It is well described by the Kroll-Ruderman term which is non-vanishing
in the chiral limit,
 \beqa
E_{0+}^{\rm thr} (\pi^+ n) &=& \, \, \, \, 
\frac{e \, g_{\pi N}}{4 \pi \sqrt{2} m \, (1
  + \mu)^{3/2}} = \, \, \, \, \, 27.6 \cdot 10^{-3}/M_{\pi} \, \, 
\, , \nonumber \\
E_{0+}^{\rm thr} (\pi^- p) &=& -\frac{e \, g_{\pi N}}{4\pi \sqrt{2} m \, (1
  + \mu)^{1/2}}= -31.7 \cdot 10^{-3}/M_{\pi}  \, \, \, ,
\label{e0let}
\eeqa
with $\mu = M_{\pi^+}/m$ and using $g_{\pi N}^2/4 \pi = 14.28$, $e^2 /
4 \pi = 1 /137.036$,
$m=928.27\,$MeV and $M_{\pi^+}= 139.57\,$MeV. In the limit $M_{\pi}
=0$, this simplifies to
\beq
E_{0+}^{\rm thr} (\pi^+ n) = -E_{0+}^{\rm thr} (\pi^- p) = 34 \cdot 
10^{-3}/M_{\pi} \quad .
\label{e0mpi0}
\eeq  
By comparing the numbers in Eq.(\ref{e0let}) and Eq.(\ref{e0mpi0}) one
notices that the kinematical corrections which are suppressed by
powers of the small parameter $\mu \simeq 1/7$ are quite substantial
for $E_{0+}^{\rm thr} (\pi^+ n)$. However, there are other corrections
which are related to pion loop diagrams and higher dimension operators.
These have been dealt with in a systematic fashion using heavy baryon 
CHPT up--to--and--including order ${\cal O}(\mu^3)$ in \cite{bkmk}. 
In this framework, one has to consider pion loop diagrams and local contact
terms whose coefficients are the LECs. These were estimated by resonance 
exchange since not enough precise data exist to pin them all down. 
Frozen kaon loops contributes to $E_{0+}^{(0)}$ and $E_{0+}^{(-)}$
while $\rho$-meson exchange contributes only to $E_{0+}^{(0)}$ and the
$\Delta$, the axial resonance and the  $c_{1-3}$ (see previous section)
 to $E_{0+}^{(-)}$, where 
$E_{0+}(\pi^+n) = \sqrt 2 (E_{0+}^{(0)} + E_{0+}^{(-)})$
and $E_{0+}(\pi^-p) = \sqrt 2 (E_{0+}^{(0)} - E_{0+}^{(-)})$. This
of course leads to some uncertainty into the result since the resonance
parameters are only known within certain ranges. Another source of uncertainty
comes 
from the regularization scale. Indeed within resonance saturation 
a spurious mild scale--dependence remains.  In the calculation $\lambda$
runs in the interval $M_\rho \le \lambda \le m_\Delta$. To ${\cal O}(\mu^3)$
one gets with $g_{\pi N} = 13.4$:
\beqa
E_{0+}^{(0)} = (-1.6 \pm 0.1) \cdot 10^{-3}/ M_\pi \,\,\, , \,\,\, 
E_{0+}^{(-)} = (21.5 \pm 0.4) \cdot 10^{-3}/ M_\pi
\label{rest}
\eeqa
The value of $g_{\pi N} $ will be discussed in the next section.
The chiral expansion is rapidly converging in contrast to neutral pion
photoproduction:
\beqa
E_{0+}^{(0)} &=& (0 -1.79 + 0.38 - 0.21) 
\cdot 10^{-3}/ M_{\pi} \,\,\, , \nonumber \\
E_{0+}^{(-)} &=& (24.01 -3.57 + 1.38 - 0.29) 
\cdot 10^{-3}/ M_{\pi}
\label{expe0}
\eeqa
where the various contributions to $E_{0+}^{(0)}$ and $E_{0+}^{(-)}$ 
of ${\cal O}(M_\pi^{n})$ with n=0,1,2,3 have been collected.
Translating these results into the
physical channels one obtains the results shown in 
the second column of table 2.

\begin{center}

\begin{tabular}{|l|c|c|c|}
    \hline
           & CHPT  &  
                  DR & 
                  Experiment  \\                      
    \hline
  $E _{0+}^{\pi^+ n}$     &  $28.2 \pm 0.6$  &  $28.0 \pm 0.2 $ & $27.9 \pm 0.5,
  28.8 \pm 0.7, 27.6 \pm 0.3 $ \\

   $E _{0+}^{\pi^- p}$ &  $-32.7 \pm 0.6 $ & $-31.7 \pm 0.2 $ & $ -31.4\pm 1.3,
  -32.2 \pm 1.2, -31.5 \pm 0.8 $\\
    \hline
  \end{tabular}

\smallskip 

\end{center}

\noindent Table 2: Predictions and data for the charged pion electric 
dipole amplitudes.

\medskip

\noindent
Also given in that table are the results of the dispersion 
theoretical (DR) analysis of \cite{dr} and the experimental ones. The first
two numbers in the last column corresponds to rather old data from 
\cite{expc1}, \cite{expc2} and  \cite{expc3} while the last ones are taken from  the recent
TRIUMF experiment on the inverse reaction $\pi^- p \to  \gamma n$ 
(see \cite{expc4}) and a preliminary SAL analysis of the reaction 
$\gamma p \to \pi^+ n$.  The overall agreement is quite good though the CHPT predictions lie
on the large side of the  most recent data. 
Clearly, we need more precise data together with a better 
knowledge of the pion nucleon coupling constant to draw a final conclusion.

With the determination of the threshold value of $E_{0+}$ for the reaction 
$\gamma n \to \pi^- p $  one can deduce the value of the difference between
the isospin 3/2 and the isospin 1/2 $\pi N$ scattering length through the
following relation based on time reversal invariance
\beq
(a_3-a_1)^2 = 9 P {q_\gamma \over q_\pi} 
E_{0+, {\rm thr}}(\gamma n \to \pi^- p) 
\label{hoeh}
\eeq
where $P$ is the experimentally well determined 
Panofsky ratio ($P= \sigma(\pi^- p \to \pi^0 n) /
  \sigma(\pi^- p \to \gamma n)  = 1.543 \pm 0.008 $)
and $q_\gamma$ and $q_\pi$ are the CM momenta of the photon and neutral
pion at the $p \pi^-$ threshold, respectively.  
In the third column of table 3, the value of the isovector
scattering length $-b_1 = (a_3-a_1)/3$ as obtained
from Eq.(\ref{hoeh}) is given. It is compared with a direct CHPT calculation of the
scattering process $\pi N \to \pi N$ (note that in that case the result is independent
of $g_{\pi N}$ to leading order, see~\cite{bkmpin}). 
The two values are consistent within
the error bars. These results clearly demonstrate the importance
of looking at different processes at the same time. One certainly needs a
better determination of the LECs, a better control on the residual scale
dependence and isospin breaking effects have to be included before one
can reach any definite conclusion.     
For comparison are given  the Karlsruhe-Helsinki (KH) result,
see \cite{hoe} and
the value obtained by \cite{psi} from the 
decay width in pionic hydrogen.

\begin{center}

\begin{tabular}{|l|c|c|c|c|}
    \hline
           & CHPT  &  
                  CHPT & 
                  KH & decay width  \\                      

    & $\pi N \to \pi N$ & $E_{0+}$ &    &  \\
    \hline
  $-b_1 [10^{-3} M_\pi^{-1}]$     &  $92 \pm 4$  &  $87.3 \pm 2.4$ & $91.3 \pm 4$
 & $96 \pm 7$ \\
     \hline
  \end{tabular}

\smallskip 

\end{center}

\noindent Table 3: Isovector $\pi N$ scattering length.  

\section{Neutral Pion Photoproduction}
\subsection{$\gamma p \to \pi^0 p$}

\begin{figure}[t]
\hskip 6.6cm
\epsfysize=6.cm
\epsffile{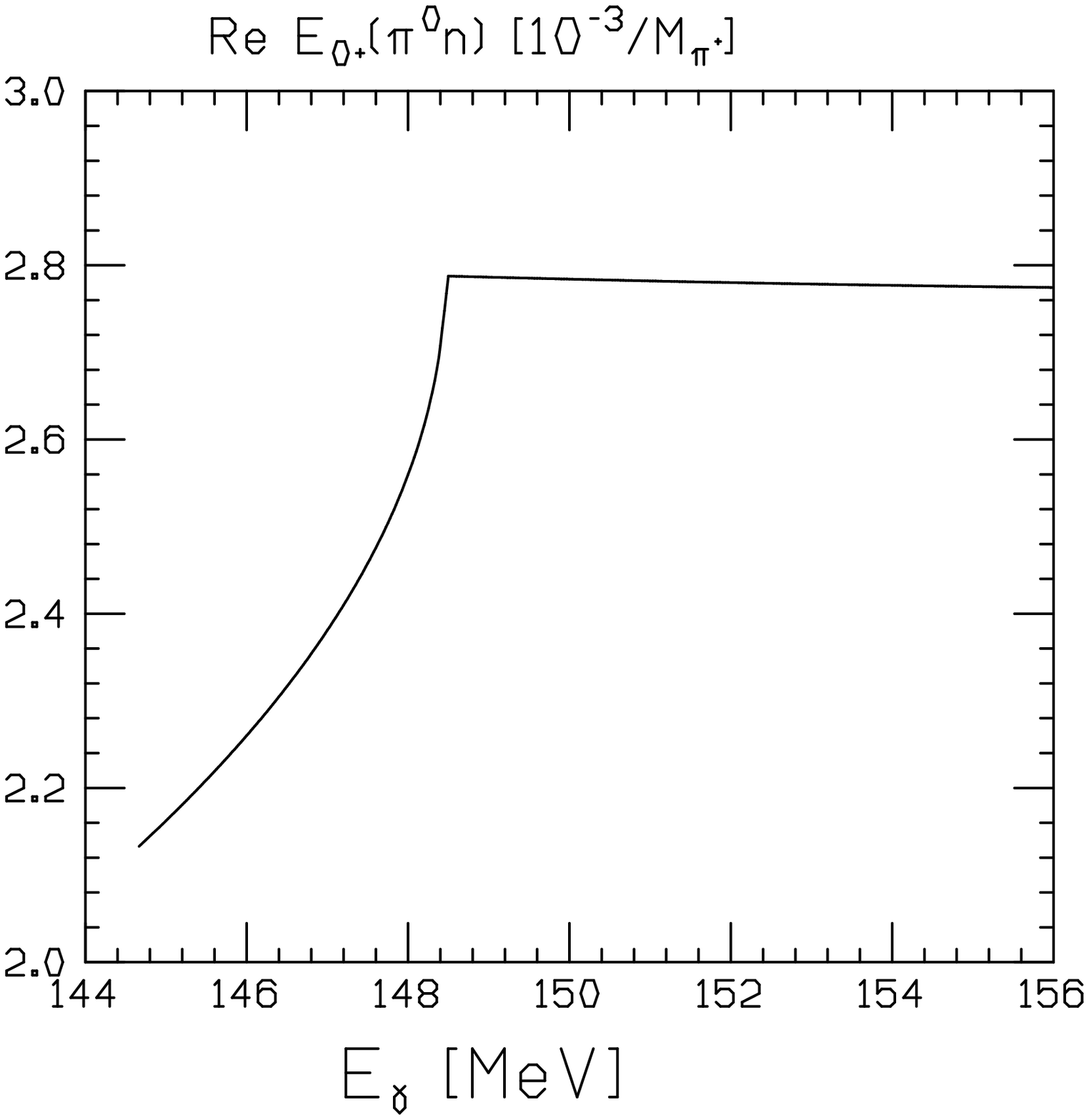}
\vskip -2.truein 
\epsfysize=4.3cm
\epsffile{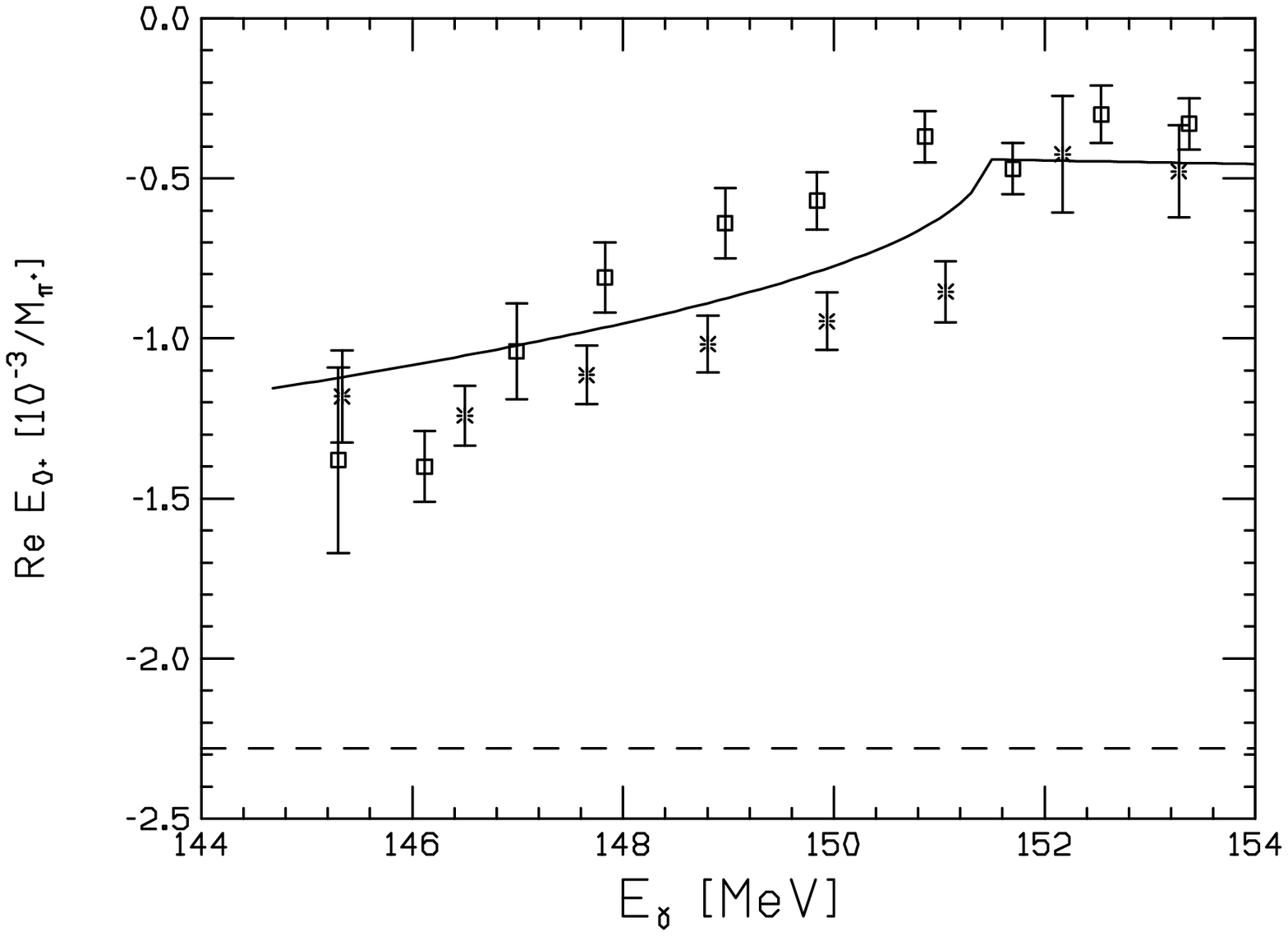}
\caption{
The real part of the electric dipole amplitude in the threshold
region. 
Left panel:$\gamma p \to \pi^0 p$.
The dashed line is the prediction of the incomplete LET and the 
data are from the multipole analysis of Fuchs et al. (1996) (boxes)
and of Bergstrom et al. (1997) (crosses).
Right panel:$\gamma n \to \pi^0 n$ }
\end{figure}

Neutral pion photoproduction has been a hot topic ever since the Saclay
and Mainz groups claimed a sizeable deviation from a so-called low 
energy theorem (LET) for the electric dipole amplitude $E_{0+}$ derived
in 1970. 
In the \cite{mit}, I reported on preliminary results which were then confirmed,
from an HBCHPT 
calculation by \cite{bkmz}. I will briefly summarize some by now well 
known facts.

$\bullet$ $E_{0+}$: To ${\cal O} (q^3)$ in HBCHPT 
two graphs contribute, the famous rescattering graph and the triangle graph.
This last important one brings in some non-analytical pieces which invalidate 
the assumptions used to derive the so-called LET of the seventies, this
one thus just being a LEG (low energy guess) (see \cite{eck:mei}).
To ${\cal O} (q^4)$
one has to take into account some more loop diagrams (1/m corrections to the
one just discussed and loops with one 
vertex or with a propagator from $L_{\pi N}^{(2)}$) plus two counterterms. 
It was shown that the expansion of $E_{0+}$ in powers of $\mu = M_\pi/m$
is slowly converging ($E_{0+}= -3.45(1-1.26+0.55)$ where the $M_\pi$, 
$M_\pi^2$ and  $M_\pi^3$ contributions are given, to be compared with
Eq.(\ref{expe0})) and therefore hard to pin down accurately. It was
concluded that at threshold this multipole cannot provide the best test
of chiral 
dynamics as was first conjectured. 

$\bullet$ P-waves: LETs have been derived for $P_{1,2}$. In contrast to
$E_{0+}$ these are very fastly converging functions of $\mu$. It is thus
particularly important to test these directly. As we will see polarization
measurements have and will be made for that purpose. It was also shown
that the P-waves are of the same chiral order as $E_{0+}$, namely they
are proportional to $|q|$ and not $ |q| |k| $ as usually postulated, 
where q and k are the pion and photon cm momenta respectively. $P_{1,2}$
have a very week energy dependence, they show no cusp effect and they
have very small imaginary parts. To ${\cal O} (q^3)$ $P_3$ is completely
dominated by a counterterm  $b_p$ which in the resonance saturation picture is
essentially described by the $\Delta$(1232) resonance . 

It turned out that new data from the TAPS collaboration (see \cite{fuchs})
were released and showed some discrepancies to the previously considered
best data of \cite{beck}. They were analysed in \cite{bkmpl}
using the same HBCHPT formalism as in 1996.
The total and differential  cross sections  were fitted and $E_{0+}$ was then 
predicted, the result is shown in Fig.1. Note that even so the convergence for
this particular observable is slow, a CHPT calculation to order $p^4$ allows
to understand its energy dependence in the threshold region. Indeed
nice agreement between theory
and experiment is obtained as confirmed by table 4 which gives the values
of $E_{0+}$ at the $\pi^0 p $ and $\pi^+ n $ threshold. For comparison is
also shown the result of the dispersion theoretical analysis (DR) of \cite{dr}.
The reaction $\gamma p \to \pi^0 p$ was also remeasured at Saskatoon, see
\cite{berg}. Between $\pi^0 p $ and $\pi^+ n $ thresholds, the SAL data
are consistent with the ones of \cite{fuchs}. For larger energies, however,
the new SAL data agree with the older Mainz data, see \cite{beck}. This does
not affect the threshold value of $E_{0+}$ but rather leads to a larger
value of $b_p$. The experimental discrepancy remains to be clarified. It 
was pointed out by R. Beck at this workshop 
that a measurement of the
so called F asymmetry in a polarized photon (circular) and polarized
target (x-direction) $\vec \gamma \vec p \to \pi^0 p$ experiment would
allow to determine Re$E_{0+}$ with a small statistical error.

\begin{center}

\begin{tabular}{|l|c|c|c|}
    \hline
           & CHPT  &  
                  DR & 
                  Experiment  \\                      
    \hline
  $E _{0+}(\pi^0 p \, {\rm thr})$     &  $-1.16$  &  $-1.22$ & $-1.31 \pm 0.8,
  -1.32 \pm 0.05 \pm 0.06 $ \\

   $E _{0+} (\pi^+ n  \,{\rm thr} )$ &  $-0.44 $ & $ \sim -0.4 $ & $ \sim -0.4 $\\
    \hline
\end{tabular}
\end{center}

\noindent Table 4: Predictions and data for the 
electric dipole amplitude for the neutral pion photoproduction off protons
at the $\pi^0 p $
and $\pi^+ n $ threshold.

\bigskip
Since the MIT workshop no real progress was made on  Im$E_{0+}$. This
is still an important piece of information missing
on the experimental side since it 
is related to the change of Re$E_{0+}$ through a dispersion relation as well
as to $\pi N$ scattering length through unitarity. A recent multipole
analysis of the Mainz $\gamma p \to \pi^0 p$ cross section measurements
by \cite{bernst}
seems to indicate some sensitivity to the assumed energy dependence of the
P-wave multipoles.   
As already
emphasized in the \cite{mit} a measurement of the polarized target asymmetry
T (T is sensitive to a linear combination of P-wave multipoles
times  Im$E_{0+}$) is awaited (see also \cite{berns} for a discussion 
of this topic). There are actually two proposals for doing a polarized 
target (y-direction) $\gamma \vec p \to  \pi^0 p$ experiment, one at
DUKE (B. Norum) and one at MAMI (A. Bernstein).

The present status of what is known from unpolarized target experiment
concerning the P-waves is summarized in table 5. There are given the three
P-waves $p_1$, $p_{23}$ with $p_{23}^2 = (p_2^2+p_3^2)/2$ and $e_{1+}$
where the small letters refer to the reduced multipoles $P_i = p_i |k| |q|$
where $P_i$ is defined in Eq.(\ref{defJ}). I kept here the standard assumed
behaviour of the P-waves though as pointed out before, it is not the proper
one. Experimental numbers 
are from \cite{fuchs} (first number for $p_1$ and $p_{23}$ ) and \cite{berg1}
(second number for $p_1$ and $p_{23}$ and number for $e_{1+}$). 
Theory and experiment agree quite nicely for $p_1$ and $p_{23}$. Note that
there is no prediction for $p_{23}$ in the case of CHPT since as pointed out
previously, 
the value of the $p_3$ mutipole mainly comes from a counterterm fitted
to experiment. Still it is important to notice that the value of $b_p$
is in excellent agreement with the resonance exchange estimate. 
In the absence of polarization data, a unique separation of the
three $P$-wave multipoles
is not possible. Thus in \cite{berg2} milder constraints provided by the 
Virginia Polytechnic Institute multipole analysis, and certain theoretical
considerations have been applied to effect the separation, thus leading to 
a value for $e_{1+}$. The one given in table 5 comes from a refined 
analysis of pion angular distributions, however the
error bar is so large that it is hard to conclude anything at present. 
Also note that the CHPT calculation was done to ${\cal O} (q^3)$ which
means that only the first term in the $\mu$ expansion of $e_{1+}$ is known.
It is certainly necessary to go one order higher to see how big the
first correction is. Work by the BKM collaboration in this direction
is in progress.  
\begin{center}

\begin{tabular}{|l|c|c|c|}
    \hline
           & CHPT  &  
                  DR & 
                  Experiment  \\                      
    \hline
  $p_1$     &  $10.33$  &  $10.52$ & $10.02 \pm 0.15,
  10.26 \pm 0.10 $ \\

   $p_{23}$ &  fit to exp & $ 10.85 $ & $ 11.44 \pm 0.09, 11.62 \pm 0.08 $ \\
   $e_1$ & $ -0.11$ & $-0.15 $ & $-0.25 \pm 0.16$ (from unpol.) \\
    \hline
\end{tabular}
\end{center}

\noindent Table 5: Predictions and data for the reduced p-waves multipoles in 
units of $|q| |k| 10^{-3} /M_{\pi^+}^3$

\medskip

What is certainly needed is a measurement of the photon asymmetry $\Sigma$ in 
a polarized photon experiment. Indeed this quantity defined as
\beq 
\Sigma(\theta) = {1 \over \epsilon_\gamma} { d \sigma^\bot -   d \sigma^\parallel  
\over d \sigma^\bot +  d \sigma^\parallel  }
\label{sig}
\eeq 
where $\epsilon_\gamma$ is the degree of linear photon 
polarization,  is proportional to $P_2$:
\beq 
\Sigma(\theta) = {1 \over \epsilon_\gamma} {q \over k} {1 \over d \sigma_0
/d \Omega }  {1 \over 2} (P_3^2 - P_2^2) {\rm sin^2 }\theta
\label{sig1}
\eeq
Thus measuring $\Sigma$ allows to test the LET for  $P_2$ and also knowing
the multipoles $P_1$ and $P_{23}$ from unpolarized target to have an unambiguous
determination of the three multipoles $M_{1+}$, $M_{1-}$ and $E_{1+}$. Data
taking was completed in Mainz by \cite{ahrens} in the threshold 
region ($E_\gamma
= 145-190$ MeV). Preliminary results were presented by 
R. Beck at this workshop.
The preliminary measured value of $\Sigma$ of ($8 \pm 5 \%$)
for $\theta = 80^\circ  - 100^\circ $ and $E_\gamma =155$MeV 
is in good agreement with the CHPT prediction of $10 \%$.

In the  calculations presented here
the $\Delta$(1232) resonance is not treated as an
explicit degree of freedom. It enters in the counterterms through the
principle of resonance saturation. It was argued by \cite{jm}
that this resonance should be explicitely taken into account since 
its mass is very close to the nucleon mass ($ 
\Delta = m_\Delta - m_N \sim 3 F_\pi$)
and the couplings of the  $N \Delta$ system to pions and
photons are very strong, e.g. $g_{\pi N \Delta} \sim 2 g_ {N N \pi}$.
Recently, \cite{hemm} proposed a systematic way of including the $\Delta$(1232)
based on an effective Lagrangian of the type ${\cal L}_{{\rm eff}} [U,N,\Delta]$
which has a systematic "small scale expansion" in terms of three small 
parameters (collectively denoted as $\epsilon$), $E_\pi/\Lambda$,
$M_\pi/\Lambda$, $\Delta/\Lambda$, with $\Lambda \in     [M_\rho,m_N,4 \pi F_\pi]$.
The method has been applied in particular to the calculation of $E_{0+}$
in neutral pion photoproduction off protons at threshold, see \cite{heme0}.  
The result obtained
is quite similar to the one of \cite{bkmpl}, meaning that
in that case no explicit $\Delta$ is needed which is quite reasonable.  
The situation might be different for the P-waves. There one indeed expect
larger sensitivity to the $\Delta$. The inclusion of the $\Delta$ as an
explicit degree of freedom for these multipoles is in progress, see \cite{hbkm}.

In all the calculations done so far the pion coupling constant $g_{\pi NN}$
was fixed to the KH value of 13.4. As pointed out by various persons, 
the chiral predictions thus obtained for $E_{0+}^{\pi^+ n}$, $E_{0+}^{\pi^+ n}$ and 
$P_1 ^{\pi^0 p}$ are consistently somewhat too big as compared to experiment.
Decreasing $g_{\pi NN}$ to $13.06 \pm 0.15$ one gets very good
agreement with experiment for this set of observables. This value turns out to 
be consistent with various recent determinations, see the proceedings of the
Seventh International Symposium on Meson-Nucleon Physics and the Structure of the
Nucleon, Vancouver (1997).
Of course,
one has to check what happens for all quantities. Furthermore, all isospin
breaking effects have to be included before one can draw any conclusion.
This is one of the major directions which remains to be explored on the 
theoretical
side. Work in this direction has started since the effective Lagrangian
including virtual photons has already been constructed by \cite{stei:mei}.

\subsection{$\gamma n \to \pi^0 n$}
As already discussed, it is very important to have predictions for the 
reaction $\gamma n \to \pi^0 n$ which is experimentally very difficult
to assess. Having fitted the LECs to the proton case \cite{bkmz} have determined 
$E_{0+} (\pi^0 n)$. A much better convergence is obtained due to the
fact that the famous triangle diagram already appears to first order which is 
already ${\cal O} (M_\pi^2)$. The first corrections are $\sim 30 \%$.   
The result is shown  on the right panel 
of Fig.1. The amplitude clearly 
exhibits the unitary cusp due to the opening of the secondary threshold
$\gamma n \to \pi^- p \to \pi^0 n$. It is sizeably larger 
in magnitude than for the proton 
(compare with the left panel).
This is  however not the case at threshold within  dispersion relations. These 
indeed tend to give values of the same size, $E_ {0+} (\pi^0 n) = 1.19$
to be compared with the CHPT result $E_ {0+} (\pi^0 n) = 2.13$ (note
that the DR treatment for the neutral channels is less stable than for the 
charged ones, see discussion in \cite{dr}). 
 
\section{Neutral pion photoproduction off the deuteron}
The elementary neutron amplitude $\gamma n \to \pi^0 n$ can only be inferred
indirectly from reactions involving few-nucleon systems like the deuteron or
$^3 {\rm He}$. \cite{bblm} concentrated on coherent neutral pion production off
deuterium in the threshold region and studied the sensitivity of the deuteron
electric dipole amplitude $E_d$ to the elementary neutron amplitude
$E_{0+}^{\pi^0 n}$. To tackle this problem it is necessary to use a framework
which allows one to systematically include and order the various contributions
arising from single and multiple scattering processes. Following the idea of
\cite{wein} one calculates matrix element of the type $<\Psi_d |{\cal K}|\Psi_d >$
by using deuteron wave functions $\Psi_d$ obtained from accurate
phenomenological NN potentials and chirally expands the kernel ${\cal K}$.
Using a large variety of these potentials allows one to assess to which 
degree of accuracy one is sensitive to the chiral symmetry constraints used
in determining the irreducible scattering kernel. 

The deuteron S-wave multipole is defined in a similar  way as the
nucleon one:
$T =  2 i \vec J \cdot \vec \epsilon E_d$
with  
$\vec J= \vec L + \vec S$ the deuteron total angular momentum.
The slope of the differential cross section at threshold takes the form:
$|\vec k| / |\vec q| d \sigma / d \Omega|_{\rm thr}
= 8/3 |E_d|^2$.
There are two types of contributions to $E_d$:

$\bullet$ {\it single scattering contribution}

The single scattering contribution is given by all diagrams where the
photon and the pion are absorbed and emitted, respectively, from one
nucleon with the second nucleon acting as a spectator (the so--called
impulse approximation), leading to
\begin{eqnarray} \label{Edss}
{E^{ss}_{d}} &=&
\frac{1+{M_\pi}/{m}} {1+{M_\pi}/{m_d}} \, \biggl\{ \frac{1}{2} \,
({E_{0+}^{{\pi ^0}p}}+{E_{0+}^{{\pi ^0}n}} ) \, \int d^{3}{p} \; 
{{\phi}}^{\ast}_{f}(\vec{p}) \, \vec{\epsilon} \cdot \vec{J} \,
{{\phi}}_{i}(\vec{p}-\vec{k}/2) \nonumber \\
&-& \frac{k}{m} \, \hat{k} \cdot \int d^{3}{p} \; \hat{p} \,
\frac{1}{2} \, ({P_{1}^{{\pi ^0}p}}+{P_{1}^{{\pi ^0}n}} ) \,
{{\phi}}^{\ast}_{f}(\vec{p})\, \vec{\epsilon} \cdot \vec{J} \,
{{\phi}}_{i}(\vec{p}-\vec{k}/2) \biggr\} \,\, ,
\end{eqnarray}
evaluated at the threshold value
$|\vec{k}| =  k_{\rm thr} 
= M_{\pi^0} - M_{\pi^0}^2 /(2\,m_d) = 130.1 \,{\rm MeV} $
and with $\vec J =  (\vec{\sigma}_1 + \vec{\sigma}_2 )/2$.
A number of remarks concerning  Eq.(\ref{Edss}) are in order. It is important
to differentiate between the $\pi^0 d$ and the $\pi^0 N$ ($N=p,n$) 
center--of--mass (COM) systems. 
At threshold in the former, the pion is produced at 
rest, it has, however, a small three--momentum in the latter, see \cite{koch}.
Consequently, one has a single--nucleon P--wave contribution proportional
to the elementary amplitudes $P_{1}^{{\pi ^0}p}$ and $P_{1}^{{\pi ^0}n}$
as defined in Eq.(\ref{defJ}). Their values, 
$P_{1}^{{\pi ^0}p} = 0.480 \, |\vec{q}\,| \,{\rm GeV}^{-2}$ and  
$P_{1}^{{\pi ^0}n} = 0.344 \, |\vec{q}\,| \,{\rm GeV}^{-2}$
have been taken from  the P--wave low--energy theorems found in \cite{bkmz}
with
$\vec{q} = \mu \, (1 - \mu ) \, \vec{p} -  \mu^2 \,
m \, (1 - 5\mu / 4 ) \, \hat{k} /2 $
($\mu = M_\pi / m$ and $\vec{p}$ is the nucleon three--momentum
in the $\pi^0 d$ COM system).
$E_d^{ss}$  has been evaluated 
using the Argonne V18, the Reid soft core (RSC),the
Nijmegen  and the Paris potential,  
${E^{ss}_{d}}=
{0.36\times {10^{-3}}}/{{{M_{\pi^+}}}}$
with an uncertainty of $\delta {E^{ss}_{d}} = {0.05\times {10^{-3}}}/
{{{M_{\pi^+}}}}$ due to the various potentials used. The P--wave contribution
amounts to a 3\% correction to the one from the S--wave, i.e. it amounts to
a minor correction. The sensitivity of the single--scattering contribution 
$E_d^{ss}$ to the elementary neutron--$\pi^0$ amplitude is given by
\begin{equation}
E_d^{ss} = 0.36 - 0.38 \cdot (2.13 - E_{0+}^{{\pi ^0}n}) \; , 
\label{sensi}
\end{equation}
all in units of $10^{-3}/M_{\pi^+}$. Consequently, for $E_{0+}^{{\pi ^0}n}
= 0$, one has $E_d^{ss} = -0.45$ which is of opposite sign to the value
based on the chiral perturbation theory prediction for $E_{0+}^{{\pi ^0}n}$. 
If one were to use the empirical value for the proton
amplitude, the single--scattering contribution would be somewhat reduced.

$\bullet$ {\it three body contribution}
 
\begin{figure}[t]
\hskip 0.2in
\epsfysize=6.0cm
\epsffile{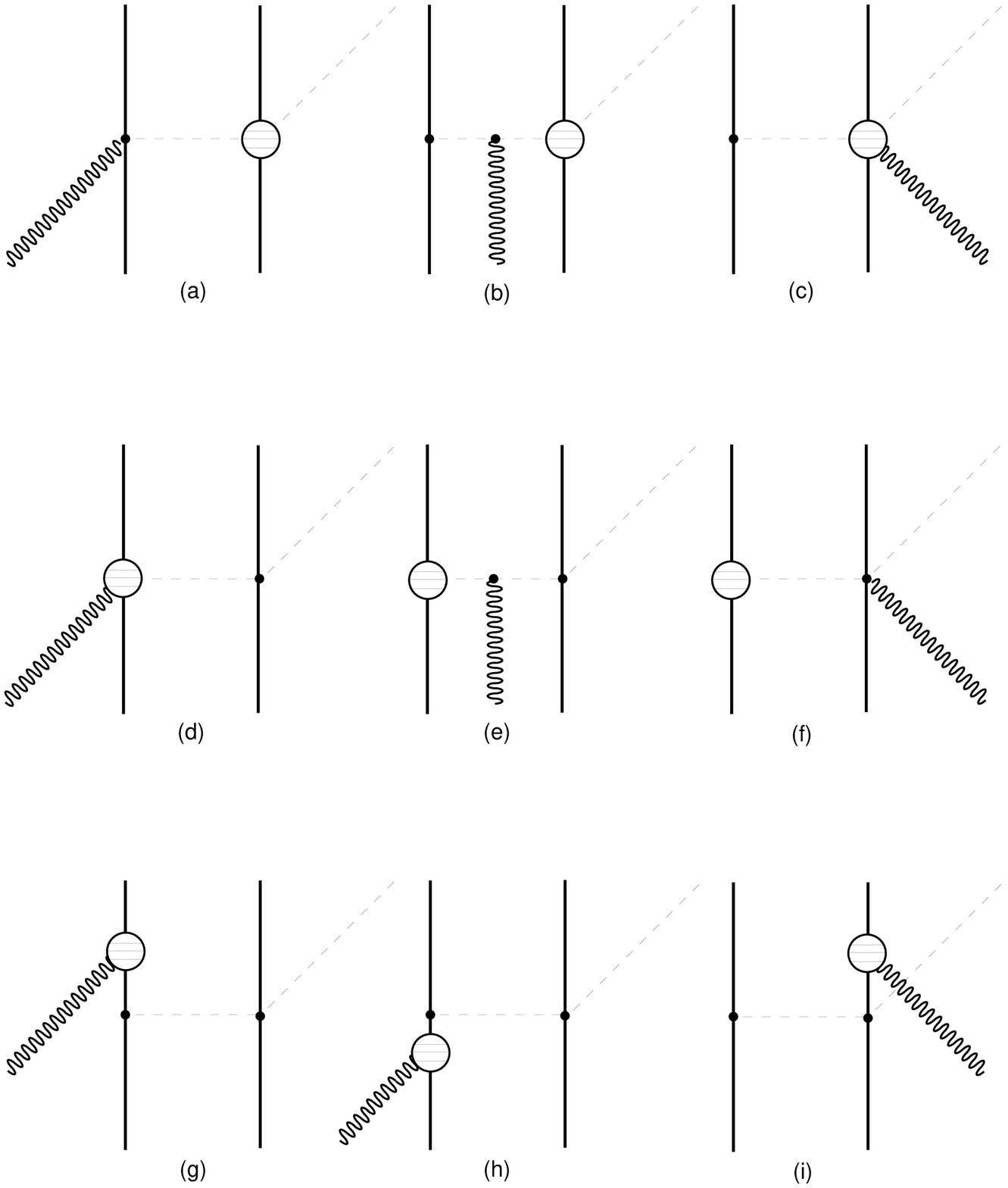}
\vskip -2.4truein
\epsfysize=6.0cm
\hspace {6cm}
\epsffile{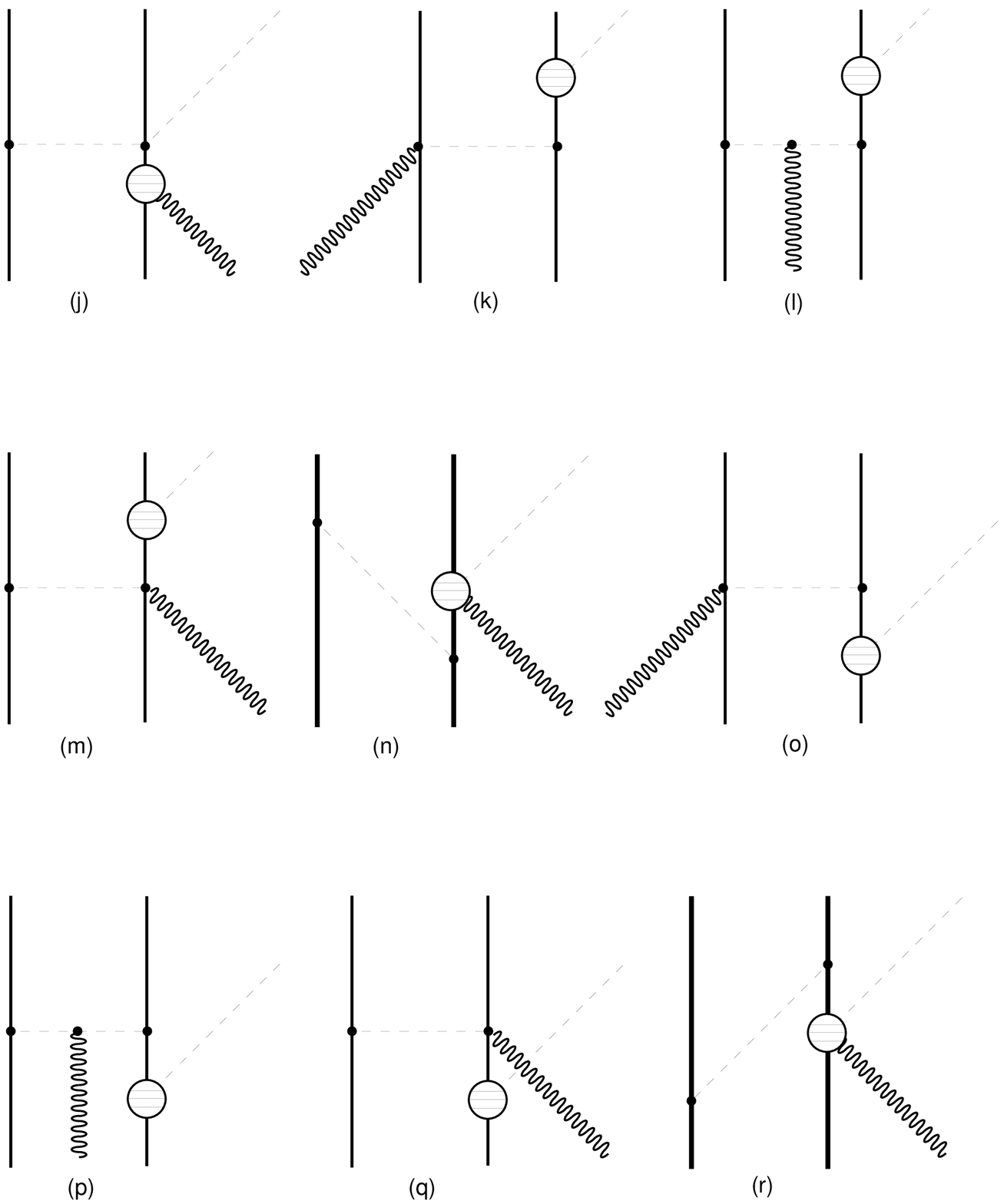}
\caption{Graphs contributing at order $q^4$ to neutral pion
photoproduction off deuteron. The circles denote an insertion
from ${\cal L}_{\pi N}^{(2)}$.} 
\end{figure}

To ${\cal O}(q^3)$ only two diagrams contribute at threshold,
see \cite{bkl}.
In one the photon couples to the pion in flight,  the other is a seagull
term which involves the charge exchange amplitude which is expected to
dominate the single scattering contribution. To  ${\cal O}(q^4)$ few
more graphs have to be added, these are shown in Fig.2. These graphs
do not involve any new unknown LEC. Furthermore there is no
contribution from a possible four-fermion contact terms. One can 
therefore calculate $E_d$ in a parameter-free manner. One finds:
\beq
E_d = E_d^{ss} + E_d^{tb,3} +  E_d^{tb,4}
= 0.36  - 1.90 - 0.25 
= -1.8 \pm 0.2
\label{red}
\eeq
in units of $10^{-3}/M_{\pi^+}$. The theoretical error is an educated guess.
As expected $E_d^{tb,3}$ is rather large.  Note that $tb,4$ is much smaller
than $tb,3$ which signals a good convergence. To see the sensitivity to the
elementary neutron $\pi^0$ amplitude, one sets the latter to zero and find
$E_d = -2.6 \times 10^{-3}/M_{\pi^+}$ which is considerably different from 
the chiral theory prediction, Eq.(\ref{red}).  
For other values of $E_{0+}^{\pi^0 n}$, $E_d$ can be calculated from
Eq.(\ref{sensi}). Obviously, the sensitivity to the neutron amplitude 
is sizeable and is not completely masked by the larger charge--exchange 
amplitude as it is often stated. On the experimental side, there exist
one determination of $E_d$, which is a reanalysis of older 
Saclay data  by \cite{arg} giving
$E_d = -1.7 \pm 0.2 \times 10^{-3}/M_{\pi^+}$ in good agreement with 
the CHPT result. This number however has to be taken with care  since
the extraction of the empirical number relies on the input from the 
elementary proton amplitude to fix a normalization constant. A more 
precise experimental determination of $E_d$ has just began. Indeed
recent data 
taking have been performed in Saskatoon with the preliminary result
$E_d = -1.45 \pm 0.09 \times 10^{-3}/M_{\pi^+}$, see these proceedings. 
This is somewhat smaller in magnitude than the CHPT value. 
The possible role of higher order unitarity corrections has been
discussed by Wilhelm at this workshop. Also 
isospin breaking effects have to be included before
one can draw any definite conclusion.
Another source of information will come from an approved  experiment at 
the Mainz Microton to measure the threshold cross section for coherent
neutral pion electroproduction off deuterium at a photon virtuality
of $k^2 = -0.075 GeV^2$. It will certainly be necessary to have a CHPT
calculation to compare with.

\section{Neutral pion electroproduction off protons}
Producing the pion with virtual photons offers further insight since one can
extract the longitudinal S-wave multipole $L_{0+}$ and also novel P-wave 
multipoles. The CHPT calculation proceeds in exactly the same way as for
photoproduction, see \cite{bkmel}.
One has, however, some new counterterms. In addition to the
expected form factor corrections, $E(k^2)=L(k^2) \propto M_\pi k^2 \delta
r_{1p}$ (where $\delta r_{1p}$ is fixed from the knowledge of the isovector
nucleon radius) one has two new LECs, $a_3$ and $a_4$ for the S-wave multipoles which are 
such that by definition $L-E \sim 1+\rho$ ($\rho =-k^2/M_\pi^2)$,
\beqa
E^{{\rm ct}} (k^2) &=& e \,M_\pi\,\{ (a_1 + a_2)\,,M_\pi^2 - a_3\, 
k^2 \} \,\,\, , \nonumber\\
L^{{\rm ct}} (k^2) &=& e \,M_\pi\,\{ (a_1 + a_2)\,,M_\pi^2 - a_3\, 
k^2 + a_4\,(M_\pi^2 -k^2)  \} \,\,\, .
\eeqa
\begin{figure}[t]
\hskip 1.5cm
\epsfysize=5.0cm
\epsffile{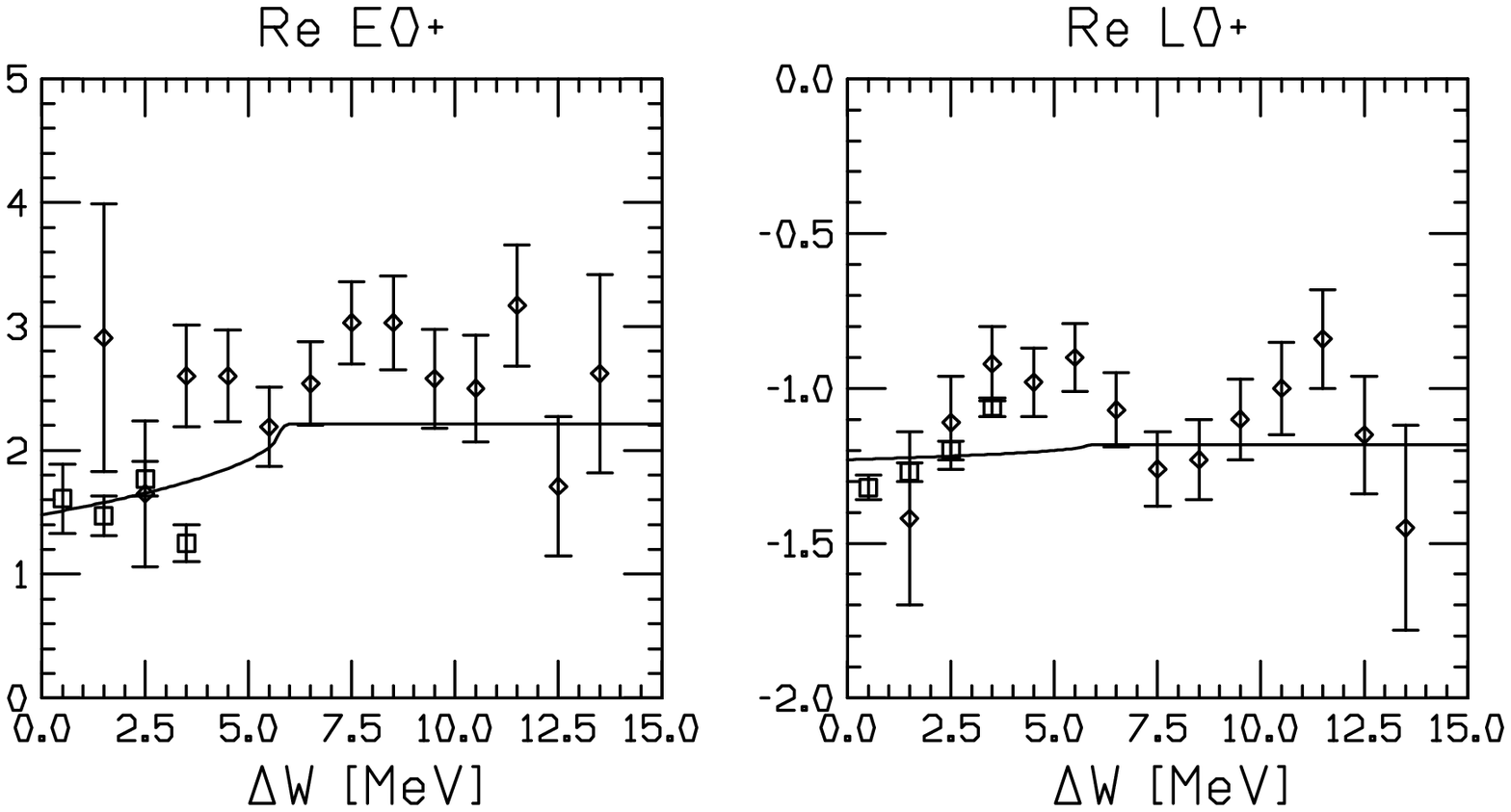}

\hspace{2cm}

\hspace{2.5cm}
\epsfysize=4.5cm
\epsffile{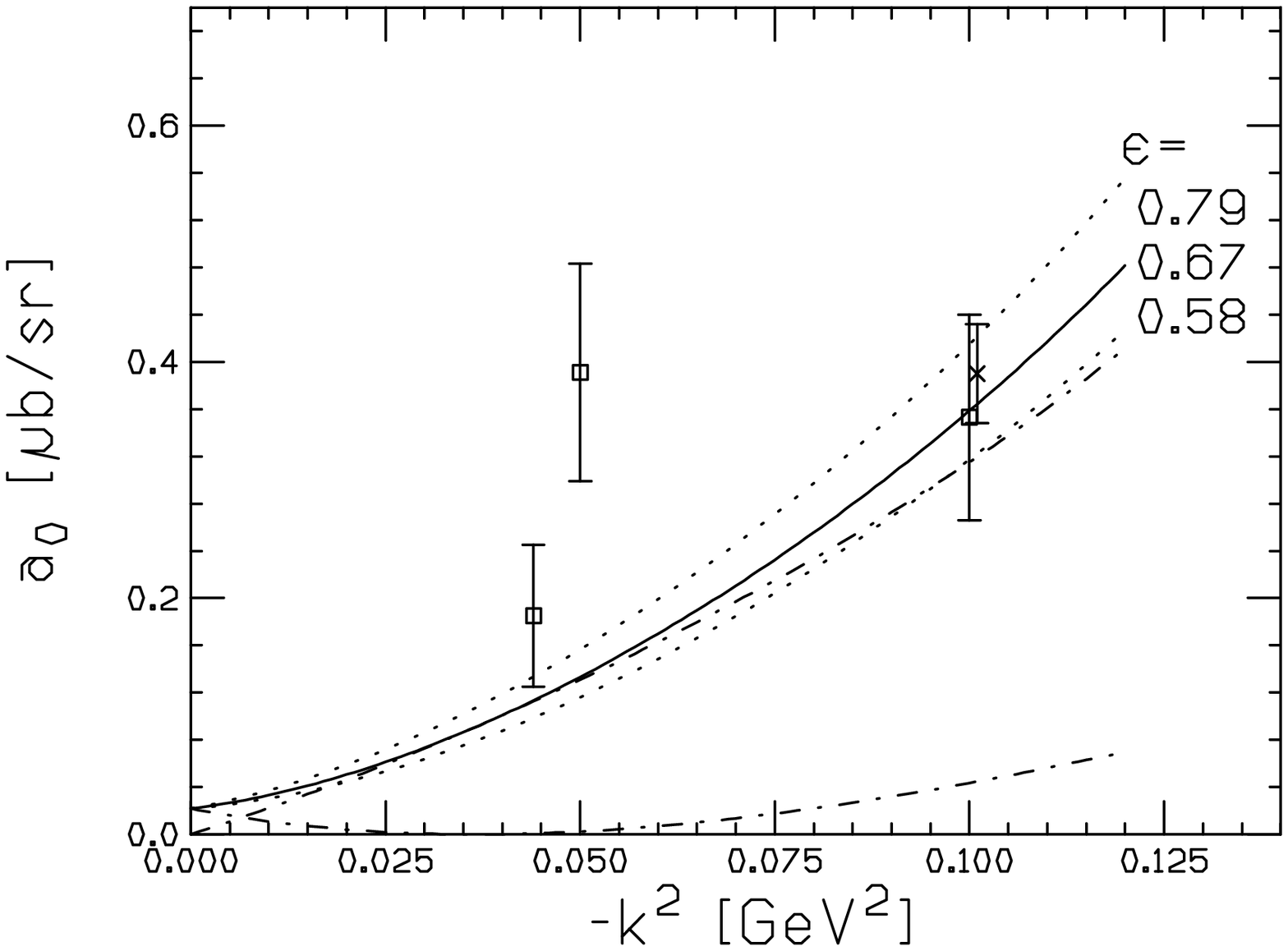}
\caption{ Upper left pannel: Re $E_{0+}$, 
upper right pannel: Re $L_{0+}$ at $k^2 = -0.1$ GeV$^2$. The diamonds
are the NIKHEF data and the boxes the Mainz data. Lower part:
S-wave cross section $a_0$ for $\epsilon = 0.67$ (solide line) in 
comparison to the data from Welch (boxes) and from van den Brink (cross). The 
dot-dashed line is the contribution from $\epsilon |L_{0+}|^2$. }

\vspace{-.5cm}

\end{figure}
It turns out that these are strongly
constrained by a soft-pion theorem, see \cite{bklm}. It implies that there
is a strong correlation between the counterterms of $E$ and $L$ to order 
$q^4$ and furthermore that $L^{{\rm ct}}$ is $k^2$ independent. 
To ${\cal O} (q^4)$ with the $k^2$-dependence of $L(k^2)$ coming solely from
the Born and loop graphs one is unable to fit the existing data. The first
corrections to the soft-pion constraint ($a_3+a_4=0$) 
away from the chiral limit
have to be included. This induces terms of the type $E_{0+}^{\rm ct} ,
L_{0+}^{\rm ct} \sim a_5 \, M_\pi^2 \, k^2$ 
which are arising from terms in the Lagrangian ${\cal L}_{\pi
  N}^{(5)}$ and are thus of higher order. These are the minimal terms
one has to take to be able to describe the data at $k^2 =
-0.1$~GeV$^2$. 
Of course, there are other counterterms at this order.
These, however, merely amount to quark mass renormalizations of the 
already considered $k^2$--independent
counter terms and have therefore been set to zero. In \cite{bkmel} a
combined fit to the NIKEHF ($\epsilon = 0.67$), see \cite{benno} and the
MAMI ($\epsilon = 0.582$ and $0.885$) , see \cite{dist}, data at $k^2 = -0.1$ 
GeV$^2$ has been performed. Re $E_{0+}$, Re $L_{0+}$ 
and the S-wave cross section $a_0 = |E_{0+}|^2 + \epsilon_L |L_{0+}|^2$
have been then 
predicted. As can be seen from Fig. 3 there is a very nice agreement
between theory and experiment. 
One notices that Re $E_{0+}$ has changed sign 
as compared to the photoproduction case, it shows the typical cups effect
at the opening of the $\pi^+ n$ threshold. In contrast, Re $L_{0+}$ is
essentially energy-independent with a very small cusp.
Note that $a_0$ is completely dominated by the 
$L_{0+}$ multipole (dot-dashed line) since $E_{0+}$ passes through zero at
$k^2 \approx - 0.04$ GeV$^2$. It should be however kept in mind that as 
one reaches values of $k^2$ of the order of $-0.1$ GeV$^2$ 
the one loop corrections are large so one should better compare at lower 
virtualities. In \cite{bkmel}, many predictions for 
$k^2 \approx - 0.05$ GeV$^2$
are given. At MAMI, data have been taken at $k^2 = - 0.05$ GeV$^2$. The 
analysis is underway, see H. Merkel, these proceedings.
Concerning the P-wave multipoles novel low--energy theorems have been
derived to ${\cal O}(q^3)$, see \cite{bkmpel}.
The combination $M_{1+}- M_1{-}$ shows a rather
good convergence. In the case of $L_{1+}$, $L_{1-}$ and $E_{1+}$, an 
${\cal O}(q^4)$ calculation is mandatory to have an idea of the next to 
leading order contributions. 
Note that these P--wave LETs have indeed been used
in the  analysis of the NIKHEF data, see~\cite{benno}.


 
\section{Conclusions}
Since the MIT workshop considerable progress has been made:

$\bullet$ charged pion photoproduction has been calculated to ${\cal O}(q^4)$,

$\bullet$ the energy dependence of $E_{0+}$ in the reaction $\gamma p \to
\pi^0 p$ in the threshold region has been understood,

$\bullet$ a study of the sensitivity of $\pi^0$ photoproduction off deuteron
to the neutron amplitude has been performed showing a rather large one,

$\bullet$ A fit to the data for $\pi^0$ electroproduction off protons at 
$k^2=-0.1$ GeV$^2$ has been performed
allowing to make predictions at smaller virtualities,

$\bullet$ another interesting process, which I had no time to discuss here,
is double neutral pion photoproduction.
It is found within an
HBCHPT calculation to ${\cal O}(q^4)$, see \cite{bkmdp},
that this photoproduction channel is significantly enhanced close to 
threshold due to pion loops. 
The experimental analysis of the TAPS data is underway, see \cite{stro}.

More precise low--energy data as well as more refined calculations are 
still needed to further test the chiral dynamics of QCD, one of the 
most important question now being related to isospin violation in the
pion-nucleon interaction and the role of electromagnetic corrections.
Work along these lines has recently started, see~\cite{stei:mei}.
%

%
%

\end{document}